\documentclass[final,5p,times,twocolumn]{elsarticle}
\usepackage{xcolor}
\usepackage{amsmath,empheq}
\usepackage{amsfonts}
\usepackage{amsthm}
\usepackage{amssymb}
\usepackage{graphicx}
\usepackage{hyperref}
\usepackage{xcolor}
\usepackage{subfigure}
\usepackage[english]{babel}

\hypersetup{
colorlinks,
linkcolor={red},
citecolor={blue},
urlcolor={blue}
}
\usepackage{color}
	\newcommand{\ncd}{\newcommand}
	\ncd{\mrm}{\mathrm}
	\ncd{\red}{\color{red}}
	\ncd{\blue}{\color{blue}}
	\ncd{\be}{\begin{equation}}
	\ncd{\ee}{\end{equation}}
 	\newcommand{\w}{{\scriptstyle \wedge}}

	\def\d{{\rm d}}

\begin{document}

\begin{frontmatter}
\title{Emission of linearly polarized photons in a strongly coupled magnetized plasma from the gauge/gravity correspondence}

\author[ciencias]{Daniel \'Avila}
\ead{davhdz06@ciencias.unam.mx}
\author[ciencias]{Tonantzin Monroy}
\ead{tona@ciencias.unam.mx}
\author[ciencias]{Francisco Nettel\fnref{corr}}
\ead{fnettel@ciencias.unam.mx}
\author[ciencias]{Leonardo Pati\~no}
\ead{leopj@ciencias.unam.mx}

\fntext[corr]{Corresponding author}

\address[ciencias]{Departamento de F\'\i sica, Facultad de Ciencias \\
	Universidad Nacional Aut\'onoma de M\'exico, A. P. 50-542, 
	M\'exico  CDMX 04510, M\'exico \\
	\vspace{0.5cm} \textit{'This article is registered under preprint number: /hep-th/2101.08802'}}

\begin{abstract}
We use holographic methods to show that photons emitted by a strongly coupled plasma subject to a magnetic field are linearly polarized regardless of their four-momentum, except when they propagate along the field direction. The gravitational dual is constructed using a 5D truncation of 10-dimensional type IIB supergravity, and includes a scalar field in addition to the constant magnetic one. In terms of the geometry of the collision experiment that we model, our statement is that any photon produced there has to be in its only polarization state parallel to the reaction plane.
\end{abstract}


\end{frontmatter}

\section{Introduction}

Thermal photons produced in the quark-gluon plasma (QGP) are very appealing probes to obtain information about the first stages of its evolution, as they are practically not scattered by the plasma \cite{Arleo:2004gn,doi:10.1146/annurev.nucl.53.041002.110533,David:2019wpt}. It was previously suggested that the QGP has global quark spin polarization in non-central heavy-ion collisions \cite{Liang:2004xn,Liang:2004ph}, and latter it was shown that this in turn leads to the polarization of the emitted photons \cite{Ipp:2007ng}, either direct \cite{Baym:2014qfa} or virtual \cite{Baym:2017gzx,Baym:2017qxy,Speranza:2018osi}. It has also been established that high energy collisions produce an intense magnetic field that points perpendicularly to the reaction plane \cite{Skokov:2009qp,Wilde:2012wc,Basar:2012bp,Andersen:2014xxa,Ayala:2018wux,Braga:2018zlu,Braga:2019yeh,Braga:2020hhs}, thus understanding its effects becomes relevant to properly analyze experimental observations. In particular, in \cite{Yee:2013qma} it was proposed that this magnetic field could lead to the quark spin polarization, and in turn induce a polarization on the emitted photons.

The gauge/gravity correspondence \cite{Maldacena:1997re} has been extensively used to explore some of the general dynamical properties of the QGP produced at high energy p-p or heavy-ion collisions, as it exists in a strongly coupled state. Although the exact gravitational dual of QCD is not known, what has been done is to consider theories similar to QCD, such as $\mathcal{N} = 4$ Super Yang-Mills (SYM) at finite temperature, and either modify them to bring them as close to QCD as possible, or use them to compute quantities that are not sensitive to the details of the theory.  A prominent example of the latter is the shear viscosity to entropy ratio, which can be computed in the strongly coupled plasma of the SYM $\mathcal{N} = 4$ theory using the gauge/gravity correspondence,  and extrapolated to the QGP produced in high energy heavy-ion collisions \cite{Policastro:2001yc}. The modifications on the other hand, have been extended to include spatial anisotropies \cite{Janik:2008tc,Mateos:2011ix,Mateos:2011tv} and, in particular, the presence of a very intense external magnetic field \cite{DHoker:2009mmn,Avila:2018hsi,Avila:2019pua,Avila:2020ved,Ammon:2020rvg}. 

The emission of photons by strongly coupled plasmas has been analyzed using different configurations in the gauge/gravity correspondence. Starting with SYM $\mathcal{N}=4$ at finite temperature in \cite{CaronHuot:2006te}, many developments have been considered to improve the modelling of the experimental context, such as a non-vanishing quark\footnote{In this context the word quark is used to refer to matter in the fundamental representation of the gauge group.} mass in the probe limit \cite{Mateos:2007yp} or in the Veneziano limit \cite{Iatrakis:2016ugz}, non-vanishing chemical potential \cite{Parnachev:2006ev,Jo:2010sg,Bu:2012zza}, spatial anisotropies \cite{Patino:2012py}, and an external magnetic field \cite{Mamo:2013efa,Arciniega:2013dqa,Wu:2013qja}. In this letter we present an extension to these alternatives. The contribution that a magnetic field turned on over the $D8/\bar{D8}$ branes of the Sakai-Sugimoto model can have to favor the production of photons with one polarization over another was explored as part of \cite{Yee:2013qma}, where the construction is not meant to include the impact that the interplay of the aforementioned field and the electromagnetic perturbations would have on the embedding of the branes that are placed in a fixed background, therefore rendering a very mild polarization.


In this letter we present a holographic analysis where the gravitational background includes the external field as part of it,  and the electromagnetic perturbations,  necessary to study photon production in the dual gauge theory,  are treated simultaneously with those of the metric to keep the solution consistent at the relevant order.  Once all these elements are assembled we can establish our main result, which is that the photons emitted by the plasma are determined to be in a strict linear polarization state, except for the degenerate case in which they propagate parallel to the magnetic field.


The holographic model we use, which was first considered in \cite{Avila:2018hsi}, is a 5-dimensional consistent truncation of 10-dimensional type IIB supergravity \cite{Cvetic:1999xp}. The magnetic field is introduced by factorizing a $U(1)$ from the $SO(6)$ symmetry of the compact space and changing it to a gauge symmetry. Thus, from the dual gauge theory perspective, the magnetic field in this case couples to the conserved current associated with a $U(1)$ subgroup of the $SU(4)$ $R-$symmetry. The importance of this model resides in the fact that it allows the introduction of massive flavor degrees of freedom by means of the embedding of D7-branes which naturally wrap a 3-cycle of the compact sector of the 10-dimensional geometry \cite{Avila:2019pua,Avila:2020ved}. However, here we will work with the unflavored 5-dimensional truncation, as this is enough to show the polarization effect caused by the magnetic field.


\section{Photon production in a strongly coupled magnetized plasma}

To calculate the photon production in the magnetized strongly coupled plasma at hand, we first establish the description in the gauge theory side. We consider a 4-dimensional $\mathcal{N} = 4$ super Yang-Mills theory over Minkowski spacetime, with gauge group $SU(N_c)$ at large $N_c$ and 'tHooft coupling $\lambda = g_{\mathrm{YM}}{}^2 N_c$. In this theory, the matter fields are in the adjoint representation of the gauge group, thus, the so-called quarks are massless. The produced photons are modeled by adding a $U(1)$ kinetic term to the SYM action that couples to the electromagnetic current associated to a $U(1)$ subgroup of the global $SU(4)$ $R$-symmetry group of the theory. Therefore, the action adopts the form of a $SU(N_c) \times U(1)$ gauge theory
	\be \label{actionSYMU1}
	S = S_{SU(N_c)} - \frac{1}{4}  \int \d^4 x \left( F^2 - 4 e A^\mu \mathcal{J}^{\rm EM}_\mu \right), 
	\ee
where $F$ is the field strength of the $U(1)$ component, comprising the background magnetic field which sources the anisotropy, $e$ stands for the electric charge, and the electromagnetic current $\mathcal{J}^{\rm{EM}}_\mu$ is given by
	\be \label{emcurrent} 
	\mathcal{J}^{\rm EM}_\mu = \bar{\Psi} \gamma_\mu \Psi + \frac{i}{2} \Phi^* (\mathcal{D}_\mu \Phi) - \frac{i}{2} (\mathcal{D}_\mu \Phi^*)^* \Phi,
	\ee
with $\Psi$ and $\Phi$ generically representing the fermionic and scalar matter content of the $SU(N_c)$ gauge theory, respectively. There is also an implicit sum over the flavor indices, and the operator $\mathcal{D}_\mu = D_\mu - i e A_\mu$ acting upon the scalar fields is the covariant derivative for the $su(N_c) \times u(1)$-connection. 

The electromagnetic coupling $\alpha_{\rm EM} = e^2/4\pi$ is small compared to 'tHooft coupling $\lambda = g_{\rm YM}{}^2 N_c$ (at large $N_c$),  so, even if the two-point correlation function necessary to compute photon production has to be calculated non-perturbatively in the $SU(N_c)$ theory that involves $\lambda$, it is enough to determine it to leading order in $\alpha_{\rm EM}$ and ignore terms of order $\mathcal{O}(\alpha_{\rm EM}^2)$. In view of the above, only the gravitational dual of the $SU(N_c)$ gauge theory is necessary, and there is no need to extend the gauge/gravity correspondence to include the full $SU(N_c) \times U(1)$ group. 

Assuming that the plasma is in thermal equilibrium when the production of the thermal photons takes place, the rate of emitted photons with null wave four-vector $k^\mu = (k^0,\vec{k})$ and polarization $\epsilon^{\mu}_{(s)}(\vec{k})$ can be calculated as \cite{CaronHuot:2006te, Mateos:2007yp}
	\begin{equation}  \label{diffproduction}
	\frac{d\Gamma_s}{d\vec{k}} = \frac{e^2}{(2\pi)^3 2 |\vec{k}|} n_B(k^0) \epsilon^{\mu}_{(s)}(\vec{k}) \epsilon^{\nu}_{(s)}(\vec{k}) \chi_{\mu\nu} (k) \bigg|_{k = 0},
	\end{equation}
where $n_B(k^0)$ is the Bose-Einstein distribution for the photon energy and the spectral density $\chi_{\mu\nu}(k) = -2\ \mathrm{Im}[G^{\rm R}_{\mu\nu}(k)]$ is given in terms of the two-point retarded correlation function for the electromagnetic current \eqref{emcurrent}
	\begin{equation} \label{2pointcorr}
	G^{\rm R}_{\mu\nu}(k) = - i \int d^4 x e^{-ik\cdot x} \Theta(t) \langle \big[ \mathcal{J}^{\rm EM}_\mu (x), \mathcal{J}^{\rm EM}_\nu(0) \big] \rangle,
	\end{equation}
with the expectation value taken in the thermal equilibrium state. 

\begin{figure}[ht!]
 \centering
 \includegraphics[width=0.5\textwidth]{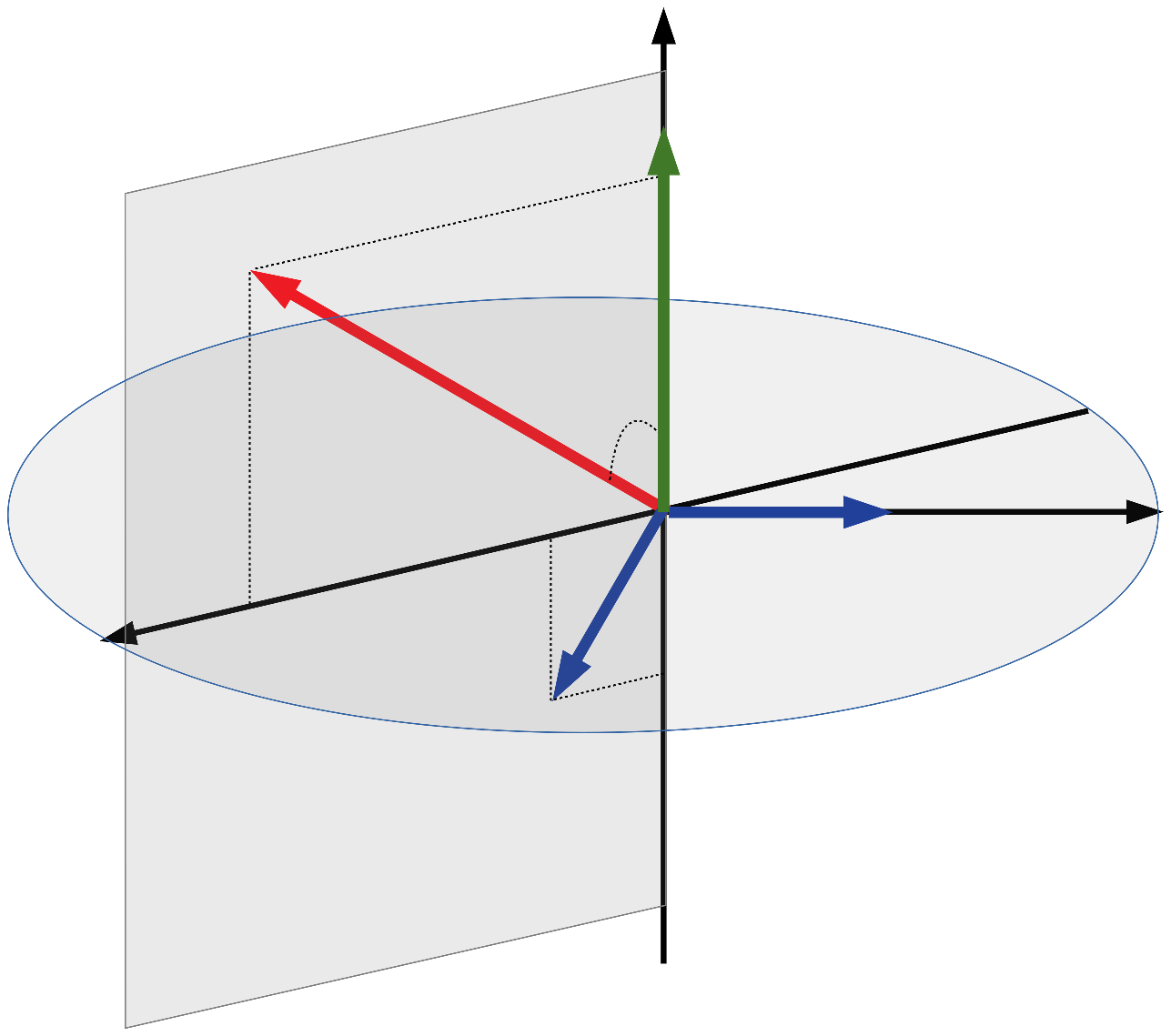}
  \put(-120,190){$z$}
  \put(-35,100){$y$}
  \put(-235,80){$x$}
  \put(-120,160){$\vec{B}$}
  \put(-205,150){$\vec{k}$}
  \put(-160,70){$\vec{\epsilon}_{(2)}$}
  \put(-100,92){$\vec{\epsilon}_{(1)}$}
\caption{{\footnotesize In this figure the spatial parts of the photon momentum $\vec{k} $ and of the polarization vectors $\vec{\epsilon}_{(1)}$ and $\vec{\epsilon}_{(2)}$ are shown. The magnetic field $\vec{B}$ points perpendicular to the reaction plane, which is depicted as a disk in the $xy$-plane. Because of the rotational symmetry around the $z$-direction, the photon momentum can be chosen to lie in the $xz$-plane without loss of generality. Our result shows that any photon produced within the plasma has to be in its only polarization state parallel to the reaction plane.}}
\label{figura}
\end{figure}

The spatial polarization four-vectors $\epsilon_{(s)}^{\mu}$ are orthogonal to the null wave four-vector with respect to the Minkowski metric, thus satisfying $\epsilon^{i}_{(s)} k^j \delta_{ij} = 0$, and can also be chosen to satisfy $\epsilon^{i}_{(1)} \epsilon^{j}_{(2)} \delta_{ij} = 0$. If we also fix our coordinate system such that the background magnetic field is directed along the $z$-direction, i.e. \mbox{$F = B\, \d x \, \w \,\d y$}, there would be a rotational symmetry on the reaction plane ($xy$-plane), allowing us to conveniently set the wave four-vector to lie in the $xz$-plane, and to denote by $\vartheta$ the angle that $\vec{k}$ forms with the background magnetic field. For these choices, the wave and polarization four-vectors take the form 
	\begin{equation} \label{wavevector}	
	k^{\mu}= k^{0}(1,\sin \vartheta,0,\cos \vartheta ),
	\end{equation}
and
	\begin{equation} \label{polvectors}
	\epsilon_{(1)}^{\mu} =(0,0,1,0), \quad  \epsilon_{(2)}^{\mu} = (0,\cos \vartheta,0,-\sin\vartheta),
	\end{equation}
respectively, and the rate of emitted photons \eqref{diffproduction} can be decomposed into 
	\begin{equation}   \label{Gammae1}
	\frac{d \Gamma_1}{d\vec{k}} \propto \chi_{yy},
	\end{equation}
for the polarization state $\epsilon_{(1)}$ and
	\begin{equation}  
	\frac{d \Gamma_2}{d\vec{k}} \propto \cos^{2}\vartheta \chi_{xx}-2\cos\vartheta\sin\vartheta \chi_{xz}+\sin^{2}\vartheta \chi_{zz}\label{Gammae2}
	\end{equation}	 
for $\epsilon_{(2)}$. An illustration of the kinematic structure is presented in Figure \ref{figura}.

\section{Gravitational dual setup}

To set up the gravitational theory dual to the gauge theory we are considering, we start from a 5-dimensional gauged supergravity theory obtained from the $S^5$ reduction of 10-dimensional type IIB supergravity and a further consistent truncation to $N=2$ supergravity theory \cite{Cvetic:1999xp}. For the purposes of this letter, it is useful to express the relevant 5-dimensional action in terms of an orthonormal frame $\{e^a\}$, $a=0,\ldots,4$, for which the dependence on the metric is explicitly shown by the Hodge dual operator $\star$ associated to it 

	\begin{equation} \label{action}
	\begin{split}
	S_{grav} &= \frac{1}{16 \pi G_5}  \int \bigg[ R_{ab}\, \w \star (e^a \w \, e^b) + \frac{4}{L^2} \sum_{i=1}^3 X_i^{-1} \star 1  \\ 
	& \qquad \qquad - \sum_{I=1}^2 \frac{1}{2}\ \d \varphi_I\, \w \star \d \varphi_I  \\ 
	& \qquad \quad - \frac{1}{2} \sum_{i=1}^3 X_i{}^{-2}\ \mathcal{F}^i \w \star \mathcal{F}^i - \mathcal{F}^1 \w \, \mathcal{F}^2 \w \, \mathcal{A}^3 \bigg] \ ,
	\end{split}
	\end{equation}
where $G_5$ is the 5-dimensional gravitational constant, $L^{-2}$ is proportional to a negative cosmological constant, $R_{ab}$ is the curvature two-form, $\varphi_I$, $I=1,2$, are two scalar fields, $\mathcal{F}^i = \d \mathcal{A}^i$, $i=1,2,3$, are the field intensities of the $U(1) \times U(1) \times U(1)$ sector (gauged field action) and
	\begin{equation}  \label{Xs}
	X_i = e^{-\tfrac{1}{2} a_i^I \varphi_I}, \quad a_i^I = (a_i^1, a_i^2) \quad \text{and} \quad \varphi_I = (\varphi_1, \varphi_2) \,
	\end{equation} 
where we are using Einstein summation convention over the $a,b,\ldots$ and $I,J , \ldots$ indices. Notice that we are using calligraphic letters for the $U(1)$ fields to distinguish them from those of the gauge theory side (cf. equation \eqref{actionSYMU1}).  

The Einstein field equation from this action is
	\begin{equation}  \label{einstein}
	\bigg[ \frac{1}{2} R^{ab} \w \, e^c \w \, e^d + \frac{1}{3! L^2} \sum_{i=1}^3 X_i{}^{-1} e^a \w \, e^b \w \, e^c \w \, e^d \bigg] \varepsilon_{abcde} = \tau_e,
	\end{equation}
where $\varepsilon_{abcde}$ is the Levi-Civita pseudo-tensor and $\tau_e$ is the energy-momentum 4-form given by the following expression
	\begin{equation} \label{emform}
	\begin{split}
	\tau_e &= \sum_{I=1}^2 \frac{1}{12} (\d\varphi_I)_a\ \varepsilon^a{}_{bcde}\ \d \varphi_{I}\ \w \, e^b \w \, e^c \w \, e^d \\
	& \hspace{1cm} - \frac{1}{2} \sum_{I=1}^2 (\d \varphi_{I})_e \star \d \varphi_I \\
	& + \frac{1}{2} \sum_{i=1}^3 X_i{}^{-2}  \bigg[ \frac{1}{4} \mathcal{F}_i{}^{ab} \varepsilon_{abcde}\ \mathcal{F}^i \w \, e^c \w \, e^d - \mathcal{F}^i_{de} \star \mathcal{F}^i \w \, e^d \bigg] \ .
	\end{split}
	\end{equation}  
The field equations for the scalar fields $\varphi_{\!I}$ are
	\begin{equation} \label{eqphi}
	\square \varphi_{I} + \frac{2}{L^2} \sum_{i=1}^3 X_i{}^{-1} a_i^I - \frac{1}{4} \sum_{i=1}^3 X_i{}^{-2} a_i^I {\mathcal{F}^i}^2 = 0\ ,
	\end{equation}
while for the Maxwell fields $\mathcal{F}^i$
	\begin{equation}  \label{eqmax}
	\d (X_i{}^{-2} \star \mathcal{F}^i) + \mathcal{F}^j \w \, \mathcal{F}^k = 0 \ ,
	\end{equation}
where $i \neq j \neq k$.
	
As can be noticed, the last expression in the action \eqref{action} is independent of the metric degrees of freedom, i.e. a topological term. 

A further truncation is imposed considering the following specific choice for the gauge and scalar fields
	\begin{equation}  \label{trunc1}
	\varphi = 2 \varphi_1 = \frac{2}{\sqrt{3}} \varphi_2, \quad \mathcal{A}^1 = 0, \quad \mathcal{A}^2 = \mathcal{A}^3 = \sqrt{2}\mathcal{A},
	\end{equation}
while
	\begin{equation}  \label{trunc2}
	a^I_1 = \left(\frac{2}{\sqrt{6}}, \sqrt{2}\right), \, a^I_2 = \left(\frac{2}{\sqrt{6}}, -\sqrt{2}\right), \, a^I_3 = \left(-\frac{4}{\sqrt{6}},0\right).
	\end{equation}
	
Einstein equation reduces in this case to
	\begin{equation} \label{einstein2}
	\begin{split}
	\bigg[ \frac{1}{2} R^{ab} & \w \, e^c \w \, e^d   \\
	&+ \frac{1}{3! L^2} \left( X^2  + 2 X^{-1}\right) e^a \w \, e^b \w \, e^c \w \, e^d \bigg] \varepsilon_{abcde} = \tau_e,
	\end{split}
	\end{equation}
with $X = e^{\frac{1}{\sqrt{6}} \varphi}$ and the energy-momentum 4-form is
	\begin{equation}  \label{emform2}
	\begin{split}
	\tau_e &= \frac{1}{12} (\d \varphi)_a\ \varepsilon^a{}_{bcde}\ \d \varphi\, \w \, e^b \w \, e^c \w \, e^d  \\ 
	& \hspace{1cm} - \frac{1}{2} (\d \varphi)_e \star \d \varphi \\
	& + \frac{1}{2} X^{-2} \mathcal{F}^{ab} \varepsilon_{abcde}\ \mathcal{F} \w \, e^c \w \, e^d - X^{-2} \mathcal{F}_{de} \star \! \mathcal{F} \, \w \, e^d.
	\end{split}
	\end{equation}

For the scalar field $\varphi$ 
	\begin{equation} \label{eqphi2}
	\square \varphi + \frac{4}{L^2} \sqrt{\frac{2}{3}} \left( X^2 - X^{-1} \right) + \sqrt{\frac{2}{3}} X^{-2} \mathcal{F}^2 = 0,
	\end{equation}
while the equations \eqref{eqmax} yield the sourceless Maxwell equations
	\begin{equation} \label{eqmax2}
	\d (X^{-2} \star \! \mathcal{F} ) = 0 ,
	\end{equation}
and a topological condition on the Maxwell field intensity
	\begin{equation}  \label{topcond}
	\mathcal{F} \, \w \, \mathcal{F} = 0.
	\end{equation}
It will be shown that this last equation entails the linear polarization effect that the strongly coupled magnetized plasma produces on the emitted photons. It is important to remark that as the field equations \eqref{einstein2}-\eqref{topcond} are derived as a consistent truncation of 10-dimensional supergravity, the condition \eqref{topcond} is unavoidable in our gravitational setup.

Equations \eqref{einstein2}-\eqref{eqmax2} can be obtained from the effective action for the gravitational configuration
	\begin{equation}  \label{effaction}
	\begin{split}
	S_{\rm eff} &= \frac{1}{16 \pi G_5}  \int \bigg[ R_{ab}\, \w \star (e^a \w \, e^b) + \frac{4}{L^2} (X^2 + 2 X^{-1})  \star 1  \\ 
	& \qquad \qquad - \frac{1}{2}\ \d \varphi \, \w \star \d \varphi  - 2 X^{-2}\ \mathcal{F} \w \star \mathcal{F}  \bigg] \ ,
	\end{split}
	\end{equation}	
while equation \eqref{topcond} must be imposed as a further constraint on the system. In the following we will take $L=1$ without loss of generality. 


In order to holographically describe the gauge theory discussed above, we need to look for solutions of the equations of motion coming from \eqref{effaction} that have a finite temperature, a magnetic field, and that are asymptotically AdS in the boundary. Such family of solutions was found in \cite{Avila:2018hsi}, where the thermodynamics of the unflavored plasma was studied extensively. The metric of any member of the family of solutions can be written as
\begin{equation}
ds^{2}=\frac{dr^{2}}{U(r)}-U(r)dt^{2}+V(r)(dx^{2}+dz^{2})+W(r)dz^{2},
\label{BGMetric}
\end{equation}
while the gauge field is taken to be a constant magnetic field pointing in the $z$-direction and the scalar field is a function of $r$ alone
\begin{equation}
\mathcal{F}=B\,dx\wedge dy, \qquad \varphi=\varphi(r).
\label{BGMagnetic}
\end{equation}
The metric asymptotes AdS$_{5}$ at the boundary located at $r=\infty$, while it features an event horizon at $r=r_{h}$ where the function $U(r)$ vanishes. Note that the constant magnetic field automatically satisfices the constraint \eqref{topcond} and the Maxwell equations for the metric \eqref{BGMetric}. The scalar field $\varphi$ is dual to a single-trace scalar operator of dimension $\Delta = 2$, and its presence has very important consequences on the thermodynamics of the plasma \cite{Avila:2018hsi,Avila:2019pua,Avila:2020ved}.

The only known analytical solution to the equations of motion is for $B=0$, while any other member of the family of solutions needs to be computed numerically. Although the general procedure is described in detail in \cite{Avila:2018hsi}, for the purpose of this letter the explicit numerical solutions are unnecessary. The polarization of the emitted photons is an analytical result. 


\section{Holographic photon production}	
	
According to the gauge/gravity correspondence, the correlation function \eqref{2pointcorr} can be obtained from a pertubative calculation in the gravitational dual theory \cite{Policastro:2002se,Policastro:2002tn,Son:2002sd,Kovtun:2005ev}. We need to consider perturbations around the background solutions \eqref{BGMetric} and \eqref{BGMagnetic}
	\begin{equation}
	g_{mn}={g^{\rm BG}}_{mn}+\epsilon\,h_{mn}
	\end{equation}
	\begin{equation} \label{efe}
	\mathcal{F} = \mathcal{F}^{\rm BG} +  \epsilon\,\d \mathcal{A}, \qquad \varphi=\varphi^{\rm BG}+\epsilon\, \phi,
	\end{equation} 
where $\epsilon$ is a small auxiliary parameter introduced to keep track of the order of the perturbations. The gauge field $\mathcal{A}$ in the gravitational side evaluated at the boundary, $\mathcal{A}^b$, is dual to the source of the gauge field $A$, i.e. to the electromagnetic current $\mathcal{J}^{\rm EM}$ in the gauge theory side, as long as we work in the gauge $\mathcal{A}_{r}=0$.

In order to obtain the correlation function \eqref{2pointcorr} we must compute the on-shell action \eqref{effaction} and take the second variation with respect to $\mathcal{A}^b$
	\begin{equation}  \label{Greengravity}
	{G^{\rm R}}^{\mu\nu}(k) = \frac{\delta^2 S_{\rm eff}^{b}}{\delta \mathcal{A}_\mu^{b} \delta \mathcal{A}_\nu^{b}}.
	\end{equation}
To follow this procedure, and given that the boundary is perpendicular to $r$, we further impose the gauge $h_{mr}=0$ for the metric perturbations (see \cite{Arciniega:2013dqa} for a discussion regarding this gauge choice). As the equations for the perturbations are coupled, in order to properly take this variation we need to resort to the methods described in \cite{Arciniega:2013dqa,Amado:2009ts,Kaminski:2009dh}. We must find in the action \eqref{effaction} those terms with second order derivatives with respect to $r$, integrate by parts to obtain the boundary term, and evaluate the latter on-shell to second order in $\epsilon$. Schematically we obtain
	\begin{equation}
	S_{\rm eff}^{b}\propto\int d^{4}x(\mathcal{O}(\mathcal{A}\mathcal{A}')+\mathcal{O}(\phi\phi')+\mathcal{O}(h^{2})+\mathcal{O}(hh')),
	\end{equation}
where the prime denotes differentiation with respect to $r$, the limit $r\rightarrow\infty$ is meant to be taken and zeroth, first and higher than second order terms in the perturbation fields are not written. Following the prescription in \cite{Arciniega:2013dqa,Kaminski:2009dh}, we see that the $\mathcal{O}(\phi\phi')$, $\mathcal{O}(h^{2})$ and $\mathcal{O}(hh')$ terms do not contribute to the second variation with respect to $\mathcal{A}^{b}$. Thus the only relevant part for this calculation is
	\begin{equation} \label{BdryAction}
	S_{\rm eff}^{b} = -\frac{1}{8\pi G_5}\int d^4 x\sqrt{-\gamma^{\rm BG}}\, U(r)^{1/2} \gamma^{\rm BG}{}^{\mu\nu} X^{-2} \,\mathcal{A}_\mu \mathcal{A}'_{\nu} ,
	\end{equation}
where ${\gamma^{\rm BG}}_{\mu\nu}$ is the background boundary metric given by
	\begin{equation}
	ds_{bdry}^{2}=-U(r)dt^{2}+V(r)(dx^{2}+dy^{2})+W(r)dz^{2}.\label{BGbds}
	\end{equation}
The only other terms that could have contributed to the second variation are of the form $\mathcal{O}(h'h')$, $\mathcal{O}(\phi'\phi')$, $\mathcal{O}(\phi'h')$, $\mathcal{O}(\mathcal{A}h')$, $\mathcal{O}(\mathcal{A}\phi')$, $\mathcal{O}(\mathcal{A}'h')$ and $\mathcal{O}(\mathcal{A}'\phi')$, but none of these appear in the action.

\section{Linearly polarized photons}

To compute the differential photon production \eqref{diffproduction} the next step is to solve the equations of motion obtained from \eqref{einstein2}-\eqref{eqmax2} for the perturbations to linear order in $\epsilon$. Given that these equations are highly non-linear, we need to resort to numerical methods to find the solutions and a detailed study of the photon production as a function of the photon momentum will be presented in a forthcoming paper. To show that the precise linear polarization of the emitted photons can be described analytically, let us turn our attention to the constraint that \eqref{topcond} imposes on their propagation. From \eqref{topcond} and \eqref{efe} we observe that up to linear order in $\epsilon$
	\begin{equation} \label{F2gen}
	\mathcal{F} \, \w \, \mathcal{F} = \mathcal{F}^{\rm BG} \, \w \, \mathcal{F}^{\rm BG} + 2\, \mathcal{F}^{\rm BG} \, \w \, \d \mathcal{A}. 
	\end{equation}
Given the orientation of the background magnetic field \eqref{BGMetric}, the first term on the right hand side of \eqref{F2gen} is zero, hence the constraint is relevant only at the order of the perturbation $\mathcal{A}$. Expressing the gauge field in the same coordinate basis of \eqref{BGMetric} and imposing the gauge $\mathcal{A}_{r}=0$ we can write
	\begin{equation} \label{Ageneral}
	\mathcal{A} =\mathcal{A}_\mu(x^\nu,r)\, \d x^\mu,
	\end{equation}
which components can be decomposed as
	\begin{equation} \label{AFourier}
	\mathcal{A}_\mu ( x^\nu,r) = \int \frac{d^4 k}{(2 \pi)^4} e^{-i k \cdot x} \tilde{\mathcal{A}}_\mu(k,r).
	\end{equation}
Thus, for an arbitrary direction of propagation \eqref{wavevector}, the constraint \eqref{topcond}, up to the relevant order in $\mathcal{A}$, reduces to
	\begin{multline} \label{topcond2}
	B\left(i\, k_0 \big(\tilde{\mathcal{A}}_z(k,r) + \mathcal{\mathcal{A}}_t(k,r) \cos \vartheta \big) \, \d t \, \w \, \d x \, \w \, \d y \, \w \, \d z  \right. \\ \left. - \tilde{\mathcal{A}}'_z(k,r)\, \d x \, \w \, \d y \, \w \, \d z \, \w \, \d r - \tilde{\mathcal{A}}'_t(k,r)\, \d t \, \w \, \d x \, \w \, \d y \, \w \, \d r \right)= 0.
	\end{multline}	
Therefore, for a non-vanishing magnetic field, the components of $\mathcal{A}$ must satisfy the following relations
	\begin{equation}  \label{constraint1}
	\tilde{\mathcal{A}}_z(k,r) - \tilde{\mathcal{A}}_t(k,r) \, \cos \vartheta= 0, 
	\end{equation}
	\begin{equation}  \label{constraint2}
	\tilde{\mathcal{A}}'_z(k,r)= 0 , \qquad  \tilde{\mathcal{A}}'_t(k,r)= 0.
	\end{equation}
In particular, \eqref{constraint2} directly implies that both $\mathcal{A}'_{t}$ and $\mathcal{A}'_{z}$ vanish,  and in turn, given the manner in which the differential equations couple the components of $\mathcal{A}$ for propagation in any direction other than $z$, this also imposes that $\mathcal{A}'_{x}=0$, reducing \eqref{BdryAction} to
	\begin{equation}  \label{RedS}
	S_{\rm eff}^{b}=-\frac{1}{8\pi G_{5}}\int d^{4}xU(r)\sqrt{W(r)} X^{-2}\, \mathcal{A}_{y}\mathcal{A}'_{y},
	\end{equation}
where \eqref{BGbds} has been substituted.

The lack of other terms in \eqref{RedS} readily shows that its substitution in \eqref{Greengravity} leads to
	\begin{equation}
	\frac{\delta^2 S_{\rm eff}^{b}}{\delta \mathcal{A}_z^{b} \delta \mathcal{A}_z^{b}} =\frac{\delta^2 S_{\rm eff}^{b}}{\delta \mathcal{A}_z^{b} \delta \mathcal{A}_x^{b}} =\frac{\delta^2 S_{\rm eff}^{b}}{\delta \mathcal{A}_x^{b} \delta \mathcal{A}_x^{b}} = 0,
	\end{equation}
and hence to the vanishing of the spectral densities $\chi_{zz}$, $\chi_{zx},$ and $\chi_{xx}$, that are the sole contributors to $\frac{d \Gamma_2}{d\vec{k}}$ in \eqref{Gammae2}.

\section{Discussion and final remarks}

We have just proved that once the back reaction to the interplay of the background magnetic field and the electromagnetic perturbations is accounted for, our model predicts that regardless of its energy, any photon produced by the plasma at an angle $\vartheta$ different from zero, will be precisely polarized along the $\epsilon_1$ direction. 

The result here presented can be stated in terms of the geometry of a collision experiment by claiming that a photon produced in such an event propagating in any direction, except that of the background magnetic field, has to be emitted in its only polarization state which is parallel to the reaction plane, that is, the radiation is linearly polarized, as ilustrated in Fig. (\ref{figura}). While the results here obtained by means of the gauge/gravity correspondence are strictly valid only for the SYM $\mathcal{N}=4$ magnetized plasma, the polarization effect on the direct photons might be a generic feature also present in those produced in the QGP, as suggested in previous works \cite{Ipp:2007ng,Baym:2014qfa,Baym:2017gzx,Baym:2017qxy,Speranza:2018osi}. It is important to stress that the present study is motivated by the evidence indicating that the QGP plasma generated in high energy experiments, like those at RHIC and the LHC, is in a strongly coupled state, thus justifying the use of the gauge/gravity correspondence to compute the photon emission rate in a strongly coupled theory. As other calculations done using this framework, ours does not provide any evidence to expect this polarization effect to persist in the weak coupling regime of QCD or other field theories.

For photons propagating parallel to the background magnetic field no restrictions are imposed; both polarization states, which are directed along the reaction plane, are present and contribute to the rate of photon production through the non-vanishing components $\frac{d \Gamma_1}{d\vec{k}}$ and $\frac{d \Gamma_2}{d\vec{k}}$. This can be seen from the field equations \eqref{einstein2}-\eqref{eqmax2} at first order in the perturbations, where setting $\vartheta = 0$ fixes $k_x=0$ through \eqref{wavevector}, and in turn implies that the restriction that otherwise set $\mathcal{A}'_x = 0$ is not longer present for this direction of propagation. In this case the term $\mathcal{A}_{x}\mathcal{A}'_{x}$ is present in \eqref{RedS} along with $\mathcal{A}_{y}\mathcal{A}'_{y}$ symmetrically.

Previous holographic studies \cite{Yee:2013qma,Arciniega:2013dqa} showed that the presence of the external magnetic field causes an increase in the production of photons with one polarization state over the other. The strict linear polarization of the photons in our model is due to the Chern-Simons term in the action \eqref{action} in the presence of a non-vanishing magnetic field, as this term is the one that gives the constriction $\mathcal{F}\wedge \mathcal{F}=0$. As previously stated, this term is unavoidable in our gravitational construction. Such term was not taken into consideration in \cite{Arciniega:2013dqa}, and while there is a Chern-Simons term in the action in \cite{Yee:2013qma}, their construction is such that the effect that the electromagnetic perturbations have on the embedding of the branes is not taken into account. While an analogous topological term could had been added to the action in \cite{Arciniega:2013dqa}, it was not compulsory in the context of that work, since its objective did not required the inclusion of internal degrees of freedom that demanded the 5-dimensional theory used therein to be a consistent truncation of 10-dimensional type IIB supergravity.\footnote{The addition of a topological term to the action in \cite{Arciniega:2013dqa} would modify Maxwell equations as $\d(\star \mathcal{F}) + \mathcal{F}\wedge\mathcal{F} = 0$ rather than appear as a independent constraint as it does in our case.}The addition of the Chern-Simons term to the action used in \cite{Arciniega:2013dqa} can be done as a separate investigation and will be considered in a future work. 

Something interesting to note is that, strictly speaking, we showed that the current density in the gauge theory \eqref{emcurrent} is polarized through \eqref{2pointcorr} in some way , as this is the operator that is dual to the gravitational gauge field $\mathcal{A}$. It is because the optical theorem that this in turn induces a polarization on the emited photons. This is reminiscent of what was proposed in \cite{Liang:2004xn,Liang:2004ph,Ipp:2007ng}, although this is not a spin polarization effect.

It would be interesting to know if considering massive quarks modifies this effect in any form. As we mentioned before, the holographic model we use allows a simple description of the embedding of flavor D7-branes \cite{Avila:2019pua,Avila:2020ved}. 

\section*{Acknowledgments}

We acknowledge partial financial support from PAPIIT IN113618, UNAM.


\bibliographystyle{ieeetr}
\bibliography{qgp-references}

\begin{thebibliography}{10}

\bibitem{Arleo:2004gn}
F.~Arleo {\em et~al.}, ``{Hard probes in heavy-ion collisions at the LHC:
  Photon physics in heavy ion collisions at the LHC},'' 11 2003.

\bibitem{doi:10.1146/annurev.nucl.53.041002.110533}
P.~Stankus, ``Direct photon production in relativistic heavy-ion collisions,''
  {\em Annual Review of Nuclear and Particle Science}, vol.~55, no.~1,
  pp.~517--554, 2005.

\bibitem{David:2019wpt}
G.~David, ``{Direct real photons in relativistic heavy ion collisions},'' {\em
  Rept. Prog. Phys.}, vol.~83, no.~4, p.~046301, 2020.

\bibitem{Liang:2004xn}
Z.-T. Liang and X.-N. Wang, ``{Spin alignment of vector mesons in non-central
  A+A collisions},'' {\em Phys. Lett. B}, vol.~629, pp.~20--26, 2005.

\bibitem{Liang:2004ph}
Z.-T. Liang and X.-N. Wang, ``{Globally polarized quark-gluon plasma in
  non-central A+A collisions},'' {\em Phys. Rev. Lett.}, vol.~94, p.~102301,
  2005.
\newblock [Erratum: Phys.Rev.Lett. 96, 039901 (2006)].

\bibitem{Ipp:2007ng}
A.~Ipp, A.~Di~Piazza, J.~Evers, and C.~H. Keitel, ``{Photon polarization as a
  probe for quark-gluon plasma dynamics},'' {\em Phys. Lett. B}, vol.~666,
  pp.~315--319, 2008.

\bibitem{Baym:2014qfa}
G.~Baym and T.~Hatsuda, ``{Polarization of Direct Photons from Gluon Anisotropy
  in Ultrarelativistic Heavy Ion Collisions},'' {\em PTEP}, vol.~2015, no.~3,
  p.~031D01, 2015.

\bibitem{Baym:2017gzx}
G.~Baym, T.~Hatsuda, and M.~Strickland, ``{Structure of virtual photon
  polarization in ultrarelativistic heavy-ion collisions},'' {\em Nucl. Phys.
  A}, vol.~967, pp.~712--715, 2017.

\bibitem{Baym:2017qxy}
G.~Baym, T.~Hatsuda, and M.~Strickland, ``{Virtual photon polarization in
  ultrarelativistic heavy-ion collisions},'' {\em Phys. Rev. C}, vol.~95,
  no.~4, p.~044907, 2017.

\bibitem{Speranza:2018osi}
E.~Speranza, A.~Jaiswal, and B.~Friman, ``{Virtual photon polarization and
  dilepton anisotropy in relativistic nucleus\textendash{}nucleus
  collisions},'' {\em Phys. Lett. B}, vol.~782, pp.~395--400, 2018.

\bibitem{Skokov:2009qp}
V.~Skokov, A.~Illarionov, and V.~Toneev, ``{Estimate of the magnetic field
  strength in heavy-ion collisions},'' {\em Int. J. Mod. Phys. A}, vol.~24,
  pp.~5925--5932, 2009.

\bibitem{Wilde:2012wc}
M.~Wilde, ``{Measurement of Direct Photons in pp and Pb-Pb Collisions with
  ALICE},'' {\em Nucl. Phys. A}, vol.~904-905, pp.~573c--576c, 2013.

\bibitem{Basar:2012bp}
G.~Basar, D.~Kharzeev, D.~Kharzeev, and V.~Skokov, ``{Conformal anomaly as a
  source of soft photons in heavy ion collisions},'' {\em Phys. Rev. Lett.},
  vol.~109, p.~202303, 2012.

\bibitem{Andersen:2014xxa}
J.~O. Andersen, W.~R. Naylor, and A.~Tranberg, ``{Phase diagram of QCD in a
  magnetic field: A review},'' {\em Rev. Mod. Phys.}, vol.~88, p.~025001, 2016.

\bibitem{Ayala:2018wux}
A.~Ayala, C.~Dominguez, S.~Hernandez-Ortiz, L.~Hernandez, M.~Loewe,
  D.~Manreza~Paret, and R.~Zamora, ``{Thermomagnetic evolution of the QCD
  strong coupling},'' {\em Phys. Rev. D}, vol.~98, no.~3, p.~031501, 2018.

\bibitem{Braga:2018zlu}
N.~R.~F. Braga and L.~F. Ferreira, ``{Heavy meson dissociation in a plasma with
  magnetic fields},'' {\em Phys. Lett. B}, vol.~783, pp.~186--192, 2018.

\bibitem{Braga:2019yeh}
N.~R.~F. Braga and L.~F. Ferreira, ``{Quasinormal modes for quarkonium in a
  plasma with magnetic fields},'' {\em Phys. Lett. B}, vol.~795, pp.~462--468,
  2019.

\bibitem{Braga:2020hhs}
N.~R.~F. Braga and R.~da~Mata, ``{Configuration entropy description of
  charmonium dissociation under the influence of magnetic fields},'' {\em Phys.
  Lett. B}, vol.~811, p.~135918, 2020.

\bibitem{Yee:2013qma}
H.-U. Yee, ``{Flows and polarization of early photons with magnetic field at
  strong coupling},'' {\em Phys. Rev. D}, vol.~88, no.~2, p.~026001, 2013.

\bibitem{Maldacena:1997re}
J.~M. Maldacena, ``{The Large N limit of superconformal field theories and
  supergravity},'' {\em Int. J. Theor. Phys.}, vol.~38, pp.~1113--1133, 1999.

\bibitem{Policastro:2001yc}
G.~Policastro, D.~T. Son, and A.~O. Starinets, ``{The Shear viscosity of
  strongly coupled N=4 supersymmetric Yang-Mills plasma},'' {\em Phys. Rev.
  Lett.}, vol.~87, p.~081601, 2001.

\bibitem{Janik:2008tc}
R.~A. Janik and P.~Witaszczyk, ``{Towards the description of anisotropic plasma
  at strong coupling},'' {\em JHEP}, vol.~09, p.~026, 2008.

\bibitem{Mateos:2011ix}
D.~Mateos and D.~Trancanelli, ``{The anisotropic N=4 super Yang-Mills plasma
  and its instabilities},'' {\em Phys. Rev. Lett.}, vol.~107, p.~101601, 2011.

\bibitem{Mateos:2011tv}
D.~Mateos and D.~Trancanelli, ``{Thermodynamics and Instabilities of a Strongly
  Coupled Anisotropic Plasma},'' {\em JHEP}, vol.~07, p.~054, 2011.

\bibitem{DHoker:2009mmn}
E.~D'Hoker and P.~Kraus, ``{Magnetic Brane Solutions in AdS},'' {\em JHEP},
  vol.~10, p.~088, 2009.

\bibitem{Avila:2018hsi}
D.~Avila and L.~Pati\~no, ``{Instability of a magnetized QGP sourced by a
  scalar operator},'' {\em JHEP}, vol.~04, p.~086, 2019.

\bibitem{Avila:2019pua}
D.~\'Avila and L.~Pati\~no, ``{Melting holographic mesons by applying a
  magnetic field},'' {\em Phys. Lett. B}, vol.~795, pp.~689--693, 2019.

\bibitem{Avila:2020ved}
D.~\'Avila and L.~Pati\~no, ``{Melting holographic mesons by cooling a
  magnetized quark gluon plasma},'' {\em JHEP}, vol.~06, p.~010, 2020.

\bibitem{Ammon:2020rvg}
M.~Ammon, S.~Grieninger, J.~Hernandez, M.~Kaminski, R.~Koirala, J.~Leiber, and
  J.~Wu, ``{Chiral hydrodynamics in strong magnetic fields},'' {\em arXiv
  preprint arXiv:2012.09183}, 2020.

\bibitem{CaronHuot:2006te}
S.~Caron-Huot, P.~Kovtun, G.~D. Moore, A.~Starinets, and L.~G. Yaffe, ``{Photon
  and dilepton production in supersymmetric Yang-Mills plasma},'' {\em JHEP},
  vol.~12, p.~015, 2006.

\bibitem{Mateos:2007yp}
D.~Mateos and L.~Patino, ``{Bright branes for strongly coupled plasmas},'' {\em
  JHEP}, vol.~11, p.~025, 2007.

\bibitem{Iatrakis:2016ugz}
I.~Iatrakis, E.~Kiritsis, C.~Shen, and D.-L. Yang, ``{Holographic Photon
  Production in Heavy Ion Collisions},'' {\em JHEP}, vol.~04, p.~035, 2017.

\bibitem{Parnachev:2006ev}
A.~Parnachev and D.~A. Sahakyan, ``{Photoemission with Chemical Potential from
  QCD Gravity Dual},'' {\em Nucl. Phys. B}, vol.~768, pp.~177--192, 2007.

\bibitem{Jo:2010sg}
K.~Jo and S.-J. Sin, ``{Photo-emission rate of sQGP at finite density},'' {\em
  Phys. Rev. D}, vol.~83, p.~026004, 2011.

\bibitem{Bu:2012zza}
Y.~Y. Bu, ``{Photoproduction and conductivity in dense holographic QCD},'' {\em
  Phys. Rev. D}, vol.~86, p.~026003, 2012.

\bibitem{Patino:2012py}
L.~Patino and D.~Trancanelli, ``{Thermal photon production in a strongly
  coupled anisotropic plasma},'' {\em JHEP}, vol.~02, p.~154, 2013.

\bibitem{Mamo:2013efa}
K.~A. Mamo, ``{Enhanced thermal photon and dilepton production in strongly
  coupled $N$ = 4 SYM plasma in strong magnetic field},'' {\em JHEP}, vol.~08,
  p.~083, 2013.

\bibitem{Arciniega:2013dqa}
G.~Arciniega, F.~Nettel, P.~Ortega, and L.~Pati\~no, ``{Brighter Branes,
  enhancement of photon production by strong magnetic fields in the
  gauge/gravity correspondence},'' {\em JHEP}, vol.~04, p.~192, 2014.

\bibitem{Wu:2013qja}
S.-Y. Wu and D.-L. Yang, ``{Holographic Photon Production with Magnetic Field
  in Anisotropic Plasmas},'' {\em JHEP}, vol.~08, p.~032, 2013.

\bibitem{Cvetic:1999xp}
M.~Cvetic, M.~Duff, P.~Hoxha, J.~T. Liu, H.~Lu, J.~Lu, R.~Martinez-Acosta,
  C.~Pope, H.~Sati, and T.~A. Tran, ``{Embedding AdS black holes in
  ten-dimensions and eleven-dimensions},'' {\em Nucl. Phys. B}, vol.~558,
  pp.~96--126, 1999.

\bibitem{Policastro:2002se}
G.~Policastro, D.~T. Son, and A.~O. Starinets, ``{From AdS / CFT correspondence
  to hydrodynamics},'' {\em JHEP}, vol.~09, p.~043, 2002.

\bibitem{Policastro:2002tn}
G.~Policastro, D.~T. Son, and A.~O. Starinets, ``{From AdS / CFT correspondence
  to hydrodynamics. 2. Sound waves},'' {\em JHEP}, vol.~12, p.~054, 2002.

\bibitem{Son:2002sd}
D.~T. Son and A.~O. Starinets, ``{Minkowski space correlators in AdS / CFT
  correspondence: Recipe and applications},'' {\em JHEP}, vol.~09, p.~042,
  2002.

\bibitem{Kovtun:2005ev}
P.~K. Kovtun and A.~O. Starinets, ``{Quasinormal modes and holography},'' {\em
  Phys. Rev. D}, vol.~72, p.~086009, 2005.

\bibitem{Amado:2009ts}
I.~Amado, M.~Kaminski, and K.~Landsteiner, ``{Hydrodynamics of Holographic
  Superconductors},'' {\em JHEP}, vol.~05, p.~021, 2009.

\bibitem{Kaminski:2009dh}
M.~Kaminski, K.~Landsteiner, J.~Mas, J.~P. Shock, and J.~Tarrio, ``{Holographic
  Operator Mixing and Quasinormal Modes on the Brane},'' {\em JHEP}, vol.~02,
  p.~021, 2010.

\end{thebibliography}

\end{document}